\documentclass[sn-mathphys-num]{sn-jnl} 

\usepackage{graphicx}
\usepackage{multirow}
\usepackage{amsmath,amssymb,amsfonts}
\usepackage{amsthm}
\usepackage{mathrsfs}
\usepackage{appendix}
\usepackage{xcolor}
\usepackage{textcomp}
\usepackage{booktabs}
\usepackage{array}
\usepackage{hyperref}
\usepackage{tabularx}
\usepackage{booktabs}
\usepackage{comment}
\usepackage{soul}       
\usepackage{ulem}       
\begin{document}

\title{Breaking the Epistemic Trap: Active Perception Under Compound Uncertainty}

\author[1]{\fnm{Chayan} \sur{Banerjee}$^*$}
\author[1]{\fnm{Ethan} \sur{Goan}$^\dagger$}

\affil[1]{\orgdiv{School of Electrical Engineering and Robotics},
\orgname{Queensland University of Technology},
\orgaddress{\city{Brisbane}, \country{Australia}}}

\renewcommand{\thefootnote}{\fnsymbol{footnote}}
\footnotetext[1]{Corresponding author: c.banerjee@qut.edu.au}
\footnotetext[2]{Contributing author: ej.goan@qut.edu.au}
\renewcommand{\thefootnote}{\arabic{footnote}}

\abstract{
Deploying reinforcement learning in safety critical domains, from autonomous vehicles to medical decision support, is constrained by failures arising when systems encounter unfamiliar conditions. We argue that the fundamental bottleneck is not individual challenges like changing dynamics or incomplete observations, but their synergistic interaction, which we term the \textit{Epistemic Trap}: agents cannot estimate their state without knowing system dynamics, nor learn dynamics without accurate state information. Proof-of-concept experiments in simulated locomotion reveal that combining these uncertainties causes failures far worse than either challenge alone, a 77\% observed degradation against the 46\% additive prediction, demonstrating that compounding failure modes can emerge and, when they do, far exceed what additive reasoning would predict. Conventional approaches typically adopt a passive epistemic stance that cannot resolve this coupled uncertainty. We propose reframing safety as an information problem. We introduce an Adaptive Safety Architecture built around three contributions. First, the Compound Uncertainty Coefficient ($\kappa$), a mutual-information based metric that quantifies how tightly state and dynamics uncertainties are coupled. Second, information-seeking policies governed by a MaxInfoRL objective that actively probe system dynamics rather than waiting for the environment to reveal itself passively. Third, regime adaptive safety constraints that tighten automatically as epistemic coupling rises. Together, these constitute a paradigm shift from passive robustness to active perception, offering a principled path toward decision making systems that operate under uncertainty, recognize their own ignorance, and act strategically to resolve it.
}

\keywords{Safe reinforcement learning, Epistemic uncertainty, Partial observability, 
Dynamics shift, Active perception, Uncertainty quantification, Bayes-adaptive POMDP}

\maketitle

\section{The Epistemic Trap: Compounding Uncertainty}

\subsection{The Fragility of Passive Control}

The use of artificial intelligence in high-stakes operational settings like clinical decision support, autonomous driving, and industrial automation has been limited largely because such systems struggle to generalize beyond the data distributions it was trained on \cite{goetz2024generalization,xu2024meta}. To address these deployment challenges, researchers have coalesced around two primary pillars: robust learning frameworks designed to handle distributional shifts \cite{pinto2017robust,qiao2025dual}, and POMDP-aware architectures that enhance state estimation to handle partial observability \cite{kaelbling1998planning,jin2025hi}.

\begin{figure}[h!]
\centering
\includegraphics[width=0.7\linewidth]{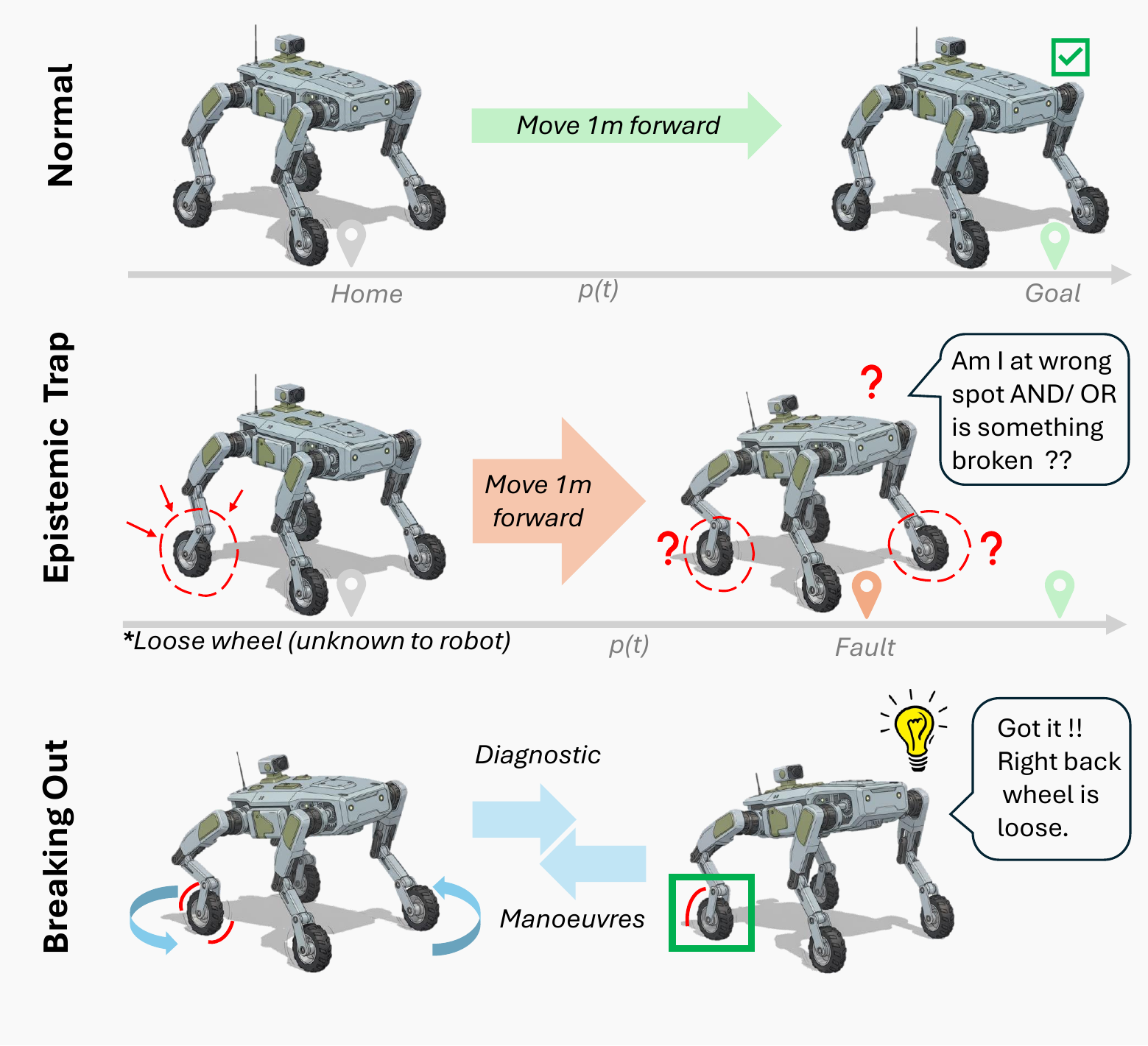}
\caption{\textbf{Visualizing the ‘Epistemic Trap’ and active resolution.}
\textit{Normal operation:} A robot navigates from Home to Goal along a planned trajectory, $p(t)$ (position over time). With reliable internal models, it executes the command accurately. \textit{The Epistemic Trap:} A hidden physical fault (loose wheel, red circle) causes the robot to drift. This creates a paralyzing dilemma: the robot cannot determine if the error is due to being in the wrong location (state uncertainty) or having broken mechanics (dynamics uncertainty), as it cannot isolate one cause without knowing the other.
\textit{Breaking out:} To resolve the ambiguity, the robot switches to active diagnostic maneuvers (blue arrows). These information-seeking actions identify the specific fault (green box), decoupling the uncertainties and restoring safe navigation.}
\label{fig:epistemic_trap}
\end{figure}

These conventional approaches share a common assumption: dynamics shifts ($\theta$) 
and partial observability (PO) can be addressed as independent challenges. This 
decomposition enables tractable solutions: standard robust RL assumes reliable observations to bound dynamics uncertainty, while classic POMDP solvers assume known transition models to resolve state ambiguity. However, this independence assumption breaks down catastrophically in deployed systems where both challenges may occur simultaneously. We refer to this simultaneous occurrence of dynamics shift and partial observability as \textit{compound uncertainty} and this condition  gives rise to the Epistemic Trap.

The fundamental issue is that state and dynamics uncertainties are not independent; they exhibit \textit{epistemic coupling}: when the agent cannot estimate state without knowing dynamics, nor learn dynamics without accurate state information creating a paralysis of reasoning (see  Figure~\ref{fig:epistemic_trap}). The illustrative example considers a quadruped wheeled robot navigating to a goal that develops a loose wheel, a fault unknown to the robot’s controller. The robot observes unexpected position errors but faces an epistemic dilemma:``Am I at the wrong location (state uncertainty)'', or ``Is my locomotion mechanism faulty (dynamics uncertainty)''. The uncertainties do not simply add, they multiply. The robot is uncertain about what it knows because it is uncertain about the rules governing how its knowledge evolves. We term this regime the \textit{Epistemic Trap}.

This circular dependency, i.e. state estimation requires known dynamics, while dynamics identification requires known state constitutes a structural failure mode that generalizes beyond robotics. Across deployment domains, failures increasingly stem from the interaction of dynamics shifts and partial observability, not from these challenges in isolation.
Self-driving cars on icy roads with heavy fog or snow experience both low-friction dynamics and impaired sensors, which together trigger unsafe maneuvers that neither challenge alone would cause \cite{knight2020snow,eidevaag2022snow}. Controlled studies of multimodal sensor fusion show that models trained under normal conditions exhibit catastrophic perception failures  in the adverse weather regimes where dynamics are most volatile \cite{bijelic2020seeing}. Autonomous surface vehicles in ice-covered waters face similar compounded risks: degraded sonar returns coincide with unpredictable ice dynamics \cite{de2023real}. Recent large-scale evaluations of vision-language-action models confirm that even foundation model-based policies suffer substantial performance drops when environmental and perceptual perturbations combine \cite{guo2025robustness}, demonstrating this challenge extends beyond traditional control domains.

Conventional Robust RL \cite{lu2024distributionally}, Safe RL \cite{wachi2024survey}, and Domain Randomization \cite{lien2023revisiting} all adopt a passive epistemic stance. These methods operate on the premise that the agent must either passively wait for the environment to reveal its true state through natural variation, or adhere to overly conservative policies derived from worst-case bounds. For instance, while recent Domain Randomization approaches attempt to mitigate over-conservatism via relaxed adversarial training \cite{lien2023revisiting}, they fundamentally remain ``blind" policies pre--hardened against uncertainty rather than capable of actively resolving it.

This passive philosophy proves inadequate in the Epistemic Trap, where natural observations provide ambiguous evidence that supports multiple incompatible hypotheses about the underlying physics. Consequently, worst-case bounding becomes a binary failure mode: it is either dangerously optimistic (if one assumes independence between state and dynamics uncertainty) or paralyzingly conservative (if one compounds worst-case scenarios for both).

In Section \ref{sec:emp_cost}, we present empirical evidence from a bipedal locomotion benchmark demonstrating that combined uncertainties trigger compounding failure modes, where performance degrades far beyond the sum of isolated errors. These non-linear collapses reveal that conventional approaches are blind to the coupling of uncertainties and are structurally limited in anticipating or preventing the failure.

Recent literature validates this concern. For instance, ActSafe \cite{as2025actsafe} demonstrates that passive exploration strategies are insufficient for safe RL, and actively seeking information about system dynamics, rather than waiting for natural variation, is necessary to guarantee safety while achieving near-optimal performance. Similarly, UNISafe \cite{seo2025unisafe} reveals that learned world models systematically `hallucinate' safety in out-of-distribution regimes when relying on passive observations alone, confirming the necessity of active epistemic probing. In the following section, we demonstrate that compound uncertainty produces measurable super-additive failures 
in controlled experiments.

\subsection{The Super-Additive Nature of Compound Failure}\label{sec:emp_cost}

The theoretical concern about compound uncertainty finds concrete support in empirical observation. We define super-additivity as a phenomenon where the total performance loss under combined stressors far exceeds the simple sum of losses from those stressors acting in isolation. While existing benchmarks typically test partial observability (POPGym \cite{morad2023popgym}) or dynamics shifts (Robust-Gymnasium \cite{gu2025robust}) in isolation, performance degrades catastrophically when these challenges combine.

\begin{figure}[htbp]
    \centering
    \includegraphics[width=0.99\linewidth]{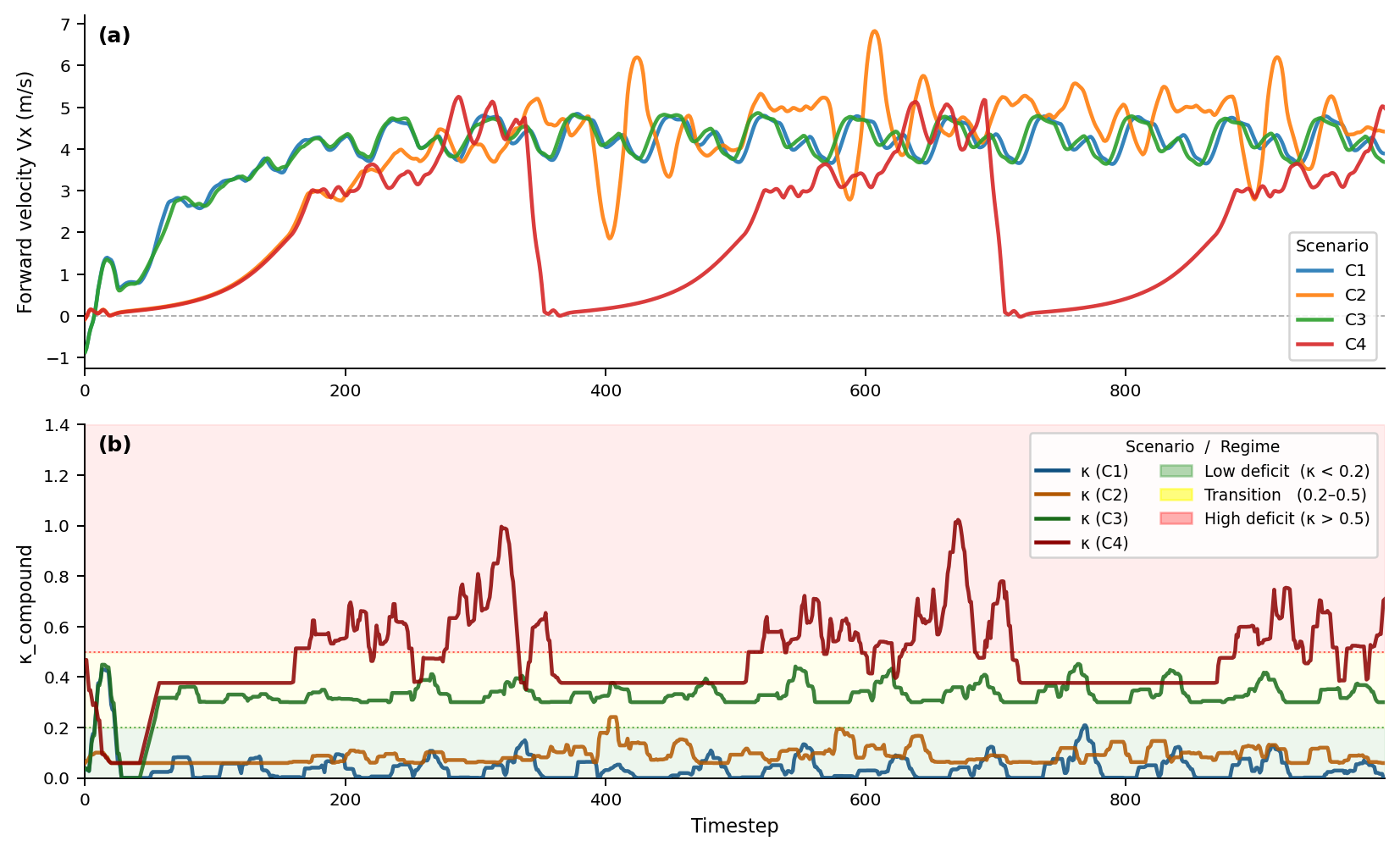}
    
    \caption{\textbf{Super-additive failure modes and epistemic coupling in bipedal locomotion.} 
    Evaluation of a Recurrent PPO agent (8M timesteps) on Walker2d-v5 across four regimes: $C_1$ (Standard), $C_2$ (Masked sensors), $C_3$ (Action delay), and $C_4$ (Combined). Perturbations introduced at $t > 50$.
    \textbf{(A) Temporal Performance:} Forward velocity ($V_x$) tracking shows that while $C_1$--$C_3$ maintain stability, $C_4$ collapses (agent falls), exhibiting super-additive loss.
    \textbf{(B) $\kappa$ as a real-time epistemic coupling signal} (formally defined in Sec:~\ref{comp_uncert}), is the Compound Uncertainty Coefficient that fires selectively under combined stressors, detecting when the agent can no longer disentangle \textit{where} it is from \textit{how} its world behaves. Post-onset means $\kappa(C_1)=0.036, \kappa(C_2)=0.089, \kappa(C_3)=0.332, \kappa(C_4)=0.496$, confirm the expected ordering $C_1 < C_2 < C_3 < C_4$. 
   \textbf{Note:} Across 120 configurations, a meaningful super-additive effect (exceeding 5\% of baseline) occurred in approximately 13\% of cases; cases with a nominally positive but marginal synergy value account for a further 12\% and are not treated as reliable instances of the phenomenon. $\text{C}_4$ shows a significant interaction effect ($p < 10^{-4}$), with an additional 413.19-unit loss beyond the sum of marginal $\text{C}_2$ and $\text{C}_3$ losses. The $\kappa$ spikes to $1.376$ ($t \approx 350, 704$) precede and signal the onset of $V_x$ collapse events in panel (A), detecting maximum epistemic surprise. Sustained high $\kappa(C_4)$ values indicate a chronic deficit that cannot be resolved passively. The legend bands correspond to the Walker2D-calibrated thresholds $\tau_{\text{low}} = 0.2$ and $\tau_{\text{high}} = 0.5$.}
    \label{fig:walker_results}
\end{figure}

\footnotetext[1]{\url{https://gymnasium.farama.org/environments/mujoco/walker2d/}}

To illustrate this compound effect, we present proof-of-concept evidence using a standard physics simulation (Walker2D)\footnotemark[1], where a virtual two-legged robot must maintain locomotion despite conflicting signals and physical changes. We evaluated a RL-agent trained under standard conditions across a spectrum of deployment scenarios: (1) standard environment, (2) sensor failure (e.g. hidden joint angles), (3) environmental shift (e.g., delayed actuator response or a heavier torso), and (4) compound uncertainty where both challenges occur simultaneously. As shown in Figure \ref{fig:walker_results}, while the agent tolerates individual stressors, the combination triggers a catastrophic collapse in forward velocity of the robot ( i.e. the robot has fallen).

Our systematic evaluation of 120 experimental configurations found that meaningful super-additive failure, defined as a compound performance loss exceeding 5\% of baseline beyond the additive prediction, occurred in approximately 13\% of cases. While infrequent, the effect was severe when it appeared: in the worst cases, compound failure exceeded the additive prediction by more than 20\% of baseline performance, reaching as much as 52\% in a single configuration. 

In a representative high-severity setting, partial observability alone reduced performance by 19\%, and dynamics shift alone by 27\%. Under an independence assumption, the expected combined degradation is approximately 46\%. Instead, performance collapsed by 77\%, indicating a super-additive failure far exceeding additive expectations. 

This non-linear degradation tends to be more frequent under temporal uncertainty (delayed actuator response), where roughly 23\% of delay-related configurations showed a meaningful super-additive effect, compared to roughly 9\% of configurations without delay. This is consistent with the idea that action delays sever the feedback loop required for cross-validation: the agent loses the ability to distinguish whether high prediction error stems from state ambiguity or dynamics mismatch.

One might assume adaptive algorithms could compensate, but current meta-learning frameworks may struggle to identify the root cause of these errors. For instance, MESA \cite{luo2021mesa} demonstrates that even sophisticated agents suffer a spike in safety violations during the initial adaptation phase. Under partial observability, this adaptation window extends significantly as the system fails to isolate the source of the error. The resulting uncertainties exhibit \textit{epistemic coupling} i.e. they do not merely add up; they couple together to distort the agent's risk estimation. This demands a structural shift: safety margins cannot remain static constraints. Instead, they must become dynamic variables, explicitly modulated by the quality of the agent's information state.

\section{Measuring the Trap: Epistemic Coupling }

\subsection{Tracking State and Dynamics Jointly}
Having established the empirical severity of compound uncertainty, we now develop a formal framework to quantify when the \textit{Epistemic Trap} becomes dangerous. 
To build intuition, consider the quadruped scenario shown in Figure~\ref{fig:epistemic_trap}. The robot must
answer two questions at the same time: \emph{``Where am I?''} and \emph{``What is the nature of the fault?''}(e.g. which wheel is loose).
The first corresponds to uncertainty over the robot's pose $(x, y, \phi)$, while the
second corresponds to uncertainty over the robot's dynamics, represented by
parameter $\theta$.

These two sources of uncertainty are tightly coupled. The robot’s belief about its
position depends on its assumptions about the wheel dynamics, for example, how much
should the robot drift when moving forward if a wheel is loose? At the same time,
inferring which wheel is faulty requires knowing the robot’s position, for example,
did the robot drift left, suggesting a fault in the right wheel?

Addressing compound uncertainty requires a rigorous mathematical foundation that treats state and dynamics uncertainty symmetrically. The B\textit{ayes-Adaptive POMDP framework} ($P_{\theta}$) \cite{ross2007bayes,duff2002optimal} provides this foundation by explicitly treating the environment dynamics parameter $\theta$ as part of the hidden state. Rather than estimating these factors in isolation, the agent must track a joint belief $b_t(s, \theta)$ at each time step $t$, simultaneously capturing its uncertainty about its location in the world and the laws governing that world. Under this unified regime, the policy must optimize three competing objectives: improving knowledge of the dynamics ($\theta$), identifying the current state ($s$), and gathering immediate reward. This formulation mathematically isolates the epistemic trap: \textit{while the problem reduces to a standard POMDP when $\theta$ is known, or to a meta-learning problem when observations are complete}, the compound uncertainty emerges specifically when both remain uncertain.

\subsection{A Metric for Compound Uncertainty} \label{comp_uncert}



To translate the theoretical concept of joint belief into a practical control signal, we require a single metric that explicitly quantifies the compound uncertainty.
Following prior work \cite{ma2023mutual, reid2025mutual, williams2010nonnegative}, we use conditional mutual information to quantify the information-theoretic coupling between state and dynamics uncertainty within the joint belief. We formalize this metric as the \emph{Compound Uncertainty Coefficient} ($\kappa$):

\begin{equation}\label{eq:kappa_compound}
\kappa  = I(s; \theta \mid b_t)
\end{equation}

Theoretically, this metric is conceptually inspired by the notion of\textit{ synergistic information} defined in the Partial Information Decomposition (PID) framework \cite{williams2010nonnegative}. 
Intuitively, $\kappa$ measures the entanglement between the two problems; where a high value signals that you cannot solve for state without also solving for dynamics, as the two are informationally inseparable.

\noindent
This formulation builds upon recent safety-critical research demonstrating the utility of mutual information (MI) under distribution shift. However, a critical distinction exists. The MISA framework leverages MI primarily for offline training regularization \cite{ma2023mutual}, and Reid et al. \cite{reid2025mutual} use it as a passive monitoring and diagnostic tool, detecting and classifying system degradation (e.g., sensor vs. actuator faults) during deployment. 

In contrast, our architecture elevates MI from a passive metric to an active, real-time control signal. By tracking the specific coupling between state and dynamics, $\kappa$ acts as a regime indicator that directly triggers corrective, information-seeking behaviors. The regime boundaries are governed by two environment-specific thresholds, $\tau_{\text{low}}$ and $\tau_{\text{high}}$, which are hyperparameters calibrated to the domain's operational risk tolerance. They partition the metric space into three operational regimes: 

\begin{itemize}
    \item Low Deficit ($\kappa < \tau_{low}$): Uncertainties act as independent noise terms, solvable by standard robust or POMDP modules.
    \item Transition ($\tau_{low} \leq \kappa \leq \tau_{high}$): Coupling is moderate. The agent balances task progress with continued uncertainty reduction.
    \item High Deficit / Epistemic Trap ($\kappa > \tau_{high}$): The transition function becomes fundamentally unidentifiable from passive observations. This signals that passive observation strategies will fail.
\end{itemize}

The thresholds $\tau_{\text{low}}$ and $\tau_{\text{high}}$ are calibrated from the observed $\kappa$ distributions of the unperturbed baseline (C1) and isolated-stressor scenarios (C2, C3), placing each condition in its expected regime; the calibration procedure is detailed in Appendix~B.

\subsubsection*{Tractable Online Approximation}
\label{sec:approx_tractable}

While $\kappa = I(s;\theta \mid b_t)$ is the ideal measure, direct computation
necessitates maintaining the full joint belief $b_t(s,\theta)$, which is
intractable in continuous systems. We therefore derive a tractable proxy. Let
$H(\cdot \mid b_t)$ denote the conditional entropy of a variable given belief
$b_t$, $H(\cdot,\cdot \mid b_t)$ its joint conditional entropy, and
$I(\cdot\,;\cdot \mid b_t)$ the conditional mutual information given $b_t$.
Applying the entropy chain rule:

\begin{equation}
I(s;\theta \mid b_t)
=
H(s \mid b_t)
+
H(\theta \mid b_t)
-
H(s,\theta \mid b_t)
\end{equation}

\noindent
Since joint conditional entropy is non-negative, $H(s,\theta \mid b_t) \geq 0$,
we obtain the provable upper bound:
\begin{equation} \label{eq:entropy_bound}
    I(s;\theta \mid b_t)
\;\leq\;
H(s \mid b_t)
+
H(\theta \mid b_t)\footnote{A detailed derivation, proxy identification,
and monotonicity argument are provided in Appendix~A.}
\end{equation}

To make this bound computable, we identify tractable proxies for each marginal
entropy term from quantities already available in the architecture. Assuming an ensemble of Gaussian predictors, $\sigma_\theta$ is a normalized
proxy for $H(\theta \mid b_t)$, while
$\sigma_s$ captures the information loss from masked state dimensions and
action delays, serving as a proxy for
$H(s \mid b_t)$. The approximation (under the assumptions of isotropic Gaussian posteriors and monotone co-variation, see detailed in Appendix~A )
\begin{equation}
    I(s;\theta \mid b_t) \;\approx\; \kappa =
\sigma_\theta + \sigma_s
    \label{eq:kappa_approx}
\end{equation}
therefore uses the entropy-additive upper bound as a tractable surrogate, where
$\sigma_\theta$ quantifies \emph{epistemic dynamics uncertainty} and $\sigma_s$
represents the \emph{state observability deficit}. We now define each component
concretely.

To compute $\sigma_\theta$, we first define the ensemble prediction error. Let
\begin{equation}
\label{eq:MSE_ensemble}
\mathrm{MSE}_t \;=\; \frac{1}{M} \sum_{m=1}^{M}
    \|f_m([\mathbf{o}_t;\,\mathbf{acc}_t],\, \mathbf{a}_t) - \Delta\mathbf{o}_t\|^2
\end{equation}
where $m \in \{1, \ldots, M\}$ indexes the ensemble members and $M$ is the
ensemble size (set to $M{=}5$ in the proof-of-concept, but treated as a
hyperparameter). Here $f_m$ denotes the $m$-th ensemble predictor,
$\mathbf{o}_t$ the observation, $\mathbf{a}_t$ the action,
$\Delta\mathbf{o}_t$ the observed kinematic delta, and $\mathbf{acc}_t$ the
acceleration feature (defined in Eq.~\ref{eq:acc} below). The normalized
excess prediction error then yields
\begin{equation}
\label{eq:sigma_theta}
\sigma_\theta \;=\;
    \frac{\operatorname{clip}\!\left(
        \dfrac{\mathrm{MSE}_t - \mu_0}{\sigma_0},\; 0,\; C
    \right)}{C}
\end{equation}
where $C$ is the clip ceiling ($C=5$), which bounds the z-score, i.e. the standardized deviation of $\mathrm{MSE}_t$ (Eq.~\ref{eq:MSE_ensemble}) from its unperturbed baseline $\mu_0$, to the interval $[0,1]$. Here $\mu_0$ and $\sigma_0$ are the mean and standard deviation of $\mathrm{MSE}$ recorded over the pre-training buffer on unperturbed transitions; dividing by the clip ceiling maps this standardized error to $[0,1]$, yielding the epistemic signal.

The $\sigma_\theta$ is derived from a bootstrapped ensemble of $M{=}5$ lightweight
two-layer MLP dynamics
predictors~\cite{osband2016deep,lakshminarayanan2017simple,chua2018deep}. Each
member is independently trained on a randomly resampled subset of baseline
transitions, inducing ensemble diversity; collective disagreement on unseen
transitions then approximates epistemic uncertainty over dynamics. Crucially,
rather than mapping raw $(\mathbf{o}_t, \mathbf{a}_t)$ directly to
$\Delta\mathbf{o}_t$, each member receives an augmented input appending the
robot's acceleration:
\begin{equation}
    \mathbf{acc}_t \;=\; \mathbf{o}_t - 2\mathbf{o}_{t-1} + \mathbf{o}_{t-2}
    \label{eq:acc}
\end{equation}
This second-order finite difference is directly disrupted by a 1-step action
delay: the phase-shifted velocity response produces a sign-inversion in
$\mathbf{acc}_t$, which remains invisible in the first-order $\Delta\mathbf{o}$
channel alone. Without this augmentation, $\sigma_\theta$ cannot distinguish
C3/C4 from C1/C2 in kinematics only observation spaces such as Walker2d-v5.

The ensemble is pre-trained on $T_{\mathrm{pre}}{=}300$ baseline transitions,
establishing a reliable prior for \emph{normal} locomotion. We then calibrate
by recording the average prediction error across the buffer, storing the mean
$\mu_0$ and standard deviation $\sigma_0$ as a \emph{noise floor}; weights are
subsequently frozen, since continued training would allow the models to ``learn''
the shifted dynamics, collapsing the error signal and masking the
fault~\cite{yu2020mopo}. During evaluation, $\sigma_\theta$ quantifies the
deviation from this noise floor as a z-score: values near zero indicate
dynamics behaving as expected, whereas positive values signify transitions the
agent could not have predicted from its baseline experience.

Turning to the second component, let
$\mathrm{PO} = \frac{d_{\mathrm{masked}}}{d_{\mathrm{total}}} \in [0,1]$
denote the fractional observation loss, where $d_{\mathrm{masked}}$ is the
number of masked observation dimensions and $d_{\mathrm{total}}$ the full
observation dimension. Let also
$\mathrm{delay} = \mathrm{clip}(\tau \cdot c_\tau,\, 0,\, 1) \in [0,1]$ be
the normalized delay penalty, where $\tau$ is the action delay in timesteps
and $c_\tau\,(=0.3)$ is an empirically calibrated scaling coefficient. The
state observability deficit is then
\begin{equation}
\label{eq:sigma_s}
\sigma_s \;=\; \mathrm{PO} + \mathrm{Delay} \cdot (1 + \mathrm{PO})
\end{equation}
where the cross-term $\mathrm{Delay} \cdot \mathrm{PO}$ encodes the
super-additive interaction, such that a delay is doubly detrimental when sensors are
already masked, since the agent lacks the observations needed to compensate
for the timing mismatch. This ensures $\sigma_s \in [0,3]$ and
$\kappa \approx \sigma_\theta + \sigma_s \in [0,4]$,
keeping the proxy bounded and meaningful as a regime indicator.

The bound in Eq.~\ref{eq:entropy_bound} is monotone, i.e.\ small when
$I(s;\theta \mid b_t)$ is small (decoupled regime) and large when coupling is
high, preserving the ordering required for reliable regime detection.
Crucially, the bound tightens in the high-coupling regime because
$H(s,\theta \mid b_t)$ is minimised precisely when $s$ and $\theta$ are
maximally entangled. This ordering is confirmed empirically:
$\kappa(C_1){=}0.036 < \kappa(C_2){=}0.089 < \kappa(C_3){=}0.332 
\kappa(C_4){=}0.496$ (Walker2d-v5), validating that the proxy captures
compound coupling without requiring full joint-belief inference. For
Walker2d-v5, empirical calibration yields $\tau_{\mathrm{low}} = 0.2$ and
$\tau_{\mathrm{high}} = 0.5$, used exclusively in the proof-of-concept
evaluation of Sec.~\ref{sec:emp_cost} and Fig.~\ref{fig:walker_results}.

\section{Breaking the Trap: Active Information-Seeking}\label{sec:break the trap}

Having quantified compound uncertainty via $\kappa $, we now present our solution to the epistemic trap: an Adaptive Safety Architecture. \textit{This architecture frames safety as an information problem by treating ($\kappa $) as a dynamic control signal rather than a static constraint.} The framework dictates a continuous transition in the agent's objective: as epistemic coupling rises ($\kappa >\tau_{\text{high}} $), the system shifts from reward maximization to Active Perception. Unlike passive observation, this regime prioritizes actions explicitly designed to decouple the joint belief $b(s,\theta)$, collapsing the high-$\kappa$ state back into a tractable regime where standard safety guarantees apply. To implement this, we must first distinguish the specific type of information seeking required.

\subsection{Seeking Dynamics, Not Just State}\label{sec:DIS}

Classical active perception in robotics predominantly focuses on State Information-Seeking \cite{bajcsy1988active}: executing actions to maximize information gain about the current state, such as maneuvering cameras to resolve occlusions or reducing localization variance. This approach is standard in Active SLAM \cite{placed2023survey} and is invaluable for resolving ambiguity when the robot's physical interaction with the world is well modeled. However, this paradigm implicitly assumes that the underlying dynamics are known or sufficiently static, a presumption that fundamentally fails when the system enters an epistemic trap. 

When a robot cannot distinguish whether a prediction error stems from a wrong position estimate or a broken actuator, standard state seeking actions (which rely on the actuator model to predict information gain) become unreliable.We must pursue a distinct strategy: \textit{Dynamics Information-Seeking (DIS)}. This involves executing exploratory actions specifically designed to isolate and identify the underlying parameters of shifted dynamics $\theta$. Unlike traditional exploration that randomly samples actions to improve reward estimates, DIS treats actions as experiments designed to probe specific physical properties. Formally, we select actions that maximize expected information gain, which highlights that the policy $\pi_{\mathrm{DIS}}$ is not maximizing reward, but rather the reduction of uncertainty regarding the dynamics parameters 
($\theta$):

\begin{equation}
\label{eq:pi_DIS}
\pi_{\mathrm{DIS}} = \arg\max_{a} \;\mathbb{E}_{o' \mid a, b_t} \left[ 
\underbrace{I\!\left(\theta; o' \mid a, b_t\right)} 
_{\substack{\text{Information gained about}\\ \text{hidden dynamics}, IG(\theta)}}
\right]
\end{equation}

This formulation frames active exploration as optimal experiment design. Rather than passively accepting data, the agent selects actions specifically to maximize the information content of future observations ($o'$), treating them as experiments to probe the causal structure of the environment. Recent implementations like ActSafe \cite{as2025actsafe} apply this principle by optimizing for epistemic uncertainty within pessimistic safe sets, achieving provable finite sample safety guarantees.

\subsection{Balancing Task Reward and Information Gain}Having established that this adaptive architecture requires DIS, we now address the challenge of incentivizing such behavior. Informative actions often conflict directly with immediate task reward,  an agent may need to pause to gather observations, execute suboptimal maneuvers to probe system responses, or maintain conservative speeds to ensure prediction accuracy. However, unlike standard intrinsic motivation approaches that prioritize novelty (\cite{pathak2017curiosity, 11288727}), our framework seeks epistemic clarity. While the pure information-seeking policy $\pi_{\mathrm{DIS}}$ defined in Eqn.~\ref{eq:pi_DIS}, captures the ideal experimental design, executing it in isolation risk preventing task completion; the agent could theoretically oscillate indefinitely to gather data without ever reaching the goal. 

Therefore, rather than applying $\pi_{\mathrm{DIS}}$ directly, we incorporate its information-theoretic signal into a composite policy, $\pi^{*}$. This redefines the exploration-exploitation dilemma: instead of a simple trade-off between reward and curiosity, exploration is repositioned as a functional requirement of the safety architecture essential for resolving the epistemic coupling that leads to system failure.
We propose the \textit{MaxInfoRL objective}, a unified control law which explicitly trades off task performance, information gain, and risk:
\begin{equation}
\label{eq:optimal_policy}
\pi^* = \arg\max_{\pi} \,
\mathbb{E}_{\pi} \Biggl[
\underbrace{R_{\text{task}}}_{\substack{\text{task} \\ \text{reward}}}
+ \alpha \cdot
\underbrace{IG(\theta)}_{\substack{\text{information gain} \\ \text{(defined in Eq.~\ref{eq:pi_DIS})}}}
- \lambda \cdot
\underbrace{R_{\text{risk}}}_{\substack{\text{safety} \\ \text{cost}}}
\Biggr]
\end{equation}

Here, $IG(\theta)$ quantifies the reduction in dynamics uncertainty, $R_{\text{risk}}$ measures the risk of constraint violation, and the coefficients $\alpha, \lambda$ modulate the competing objectives. Crucially, $\alpha = \alpha(\kappa )$ is a monotonically increasing function of epistemic coupling, consistent with the regime transitions in Table~\ref{tab:regimes}: it approaches zero in the Low Deficit regime ($\kappa < \tau_{\text{low}}$, restoring pure task optimization) and dominates in the High Deficit regime ($\kappa > \tau_{\text{high}}$), making information gain the primary objective. This treats information gain as a long-term safety value resolving $\theta$ uncertainty at time $t$ prevents future violations caused by ignorance. Empirical and theoretical evidence supports this approach: studies of intrinsic motivation show that explicit information-seeking substantially improves sample efficiency and adaptation \cite{as2025actsafe,carrasco2025uncertainty}, and recent work demonstrates that information-directed exploration bonuses can complement and outperform count-based methods in structured contextual setting \cite{henaff2023study}.

\subsection{Adaptive Safety: The Complete Framework}

The MaxInfoRL objective provides the \emph{what}, i.e. a unified control law balancing task reward, information gain, and risk. We now address the \emph{how}, i.e. a closed-loop architecture that continuously adapts safety margins based on the agent's evolving information state.
The core insight is that safety constraints should not be static thresholds, but dynamic variables tied to epistemic coupling. When $\kappa$ is high, the agent cannot reliably predict the consequences of its actions and must therefore operate conservatively while actively seeking clarifying information. As uncertainty resolves, constraints can relax, restoring performance. This yields an adaptive constraint formulation:
\begin{equation}\label{eq:risk}
\max_\pi \mathbb{E}_\pi[R_{\text{task}}] \quad \text{subject to:} \quad \mathbb{E}_\pi[R_{\text{risk}}] \leq \delta(\kappa )
\end{equation}
where the safety budget $\delta(\cdot)$ tightens as epistemic coupling rises. Recent empirical work validates this philosophy, ConstrainedZero \cite{moss2024constrainedzero} dynamically tightens safety bounds based on probabilistic failure estimates during planning, while UNISafe \cite{seo2025unisafe} uses conformal prediction to trigger fallback policies when uncertainty exceeds calibrated thresholds.

\begin{figure}[htbp]
\centering
\includegraphics[width=0.7\linewidth]{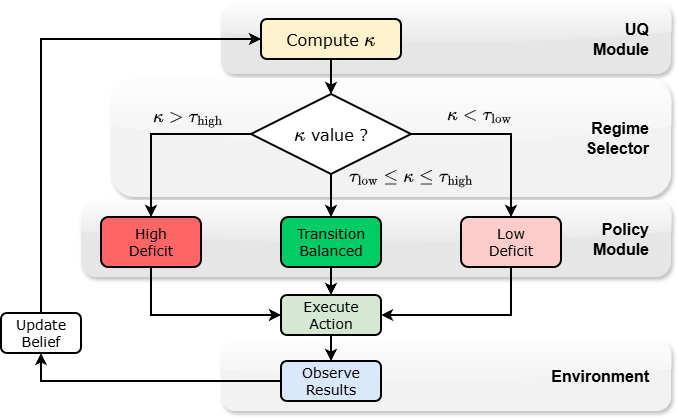}
\caption{The uncertainty quantification (UQ) module continuously estimates the \textbf{compound uncertainty coefficient} from the agent's joint belief $b_{t}(s,\theta)$. This value determines the operational regime: when uncertainty coupling is high, the agent prioritizes information-seeking diagnostic actions with conservative safety margins; during transition, the agent balances task progress with continued uncertainty reduction; when coupling is resolved, the agent maximizes task performance. Actions generate observations that update beliefs, completing the feedback loop. See Table~\ref{tab:regimes} for regime-specific behaviors.
}
\label{fig:control_loop}
\end{figure}

Figure~\ref{fig:control_loop} illustrates the complete control loop. The \textit{Uncertainty Quantification} (UQ) module continuously estimates $\kappa$ from the joint belief $b(s,\theta)$. This estimate feeds into a \textit{Regime Selector} that determines the agent's operational mode. This modulates the information-seeking weight $\alpha$ and adjusts the safety penalty $\lambda$ to enforce the tightening constraint $\delta(\kappa )$ (Eq.~\ref{eq:risk}). Crucially, the \textit{Policy Module} optimizes the MaxInfoRL objective (Eq.~\ref{eq:optimal_policy}) using these updated weights, generating actions that update the belief and close the loop.

\newcolumntype{Y}{>{\raggedright\arraybackslash}X}

\begin{table}[htbp]
\caption{Regime-Specific Policy Behavior.}
\label{tab:regimes}
\centering
\small
\begin{tabularx}{\textwidth}{>{\raggedright\arraybackslash}p{2cm} 
                             >{\raggedright\arraybackslash}p{2.5cm} 
                             >{\raggedright\arraybackslash}p{3cm} 
                             Y}
\toprule
\textbf{Regime} & \textbf{$\kappa$ Value} & \textbf{Primary Objective} & \textbf{Example Behavior (Autonomous Vehicle)} \\
\midrule
High Deficit & $\kappa > \tau_\text{high}$ (Strong coupling) & Information Gain (DIS) & Halt or perform diagnostic gaits to isolate faults, or revert to safe limp mode \\
Transition & $\tau_\text{low} \le \kappa \le \tau_\text{high}$ (Moderate coupling) & Balanced $R_{task} + \alpha \cdot IG$ & Cautious navigation with periodic diagnostic checks to refine fault hypothesis \\
Low Deficit & $\kappa < \tau_\text{low}$ (Resolved) & Maximize Task Reward & Navigate efficiently using optimal gait with confidence in wheel status \\
\bottomrule
\end{tabularx}
\end{table}

Returning to our quadruped example (Figure~\ref{fig:epistemic_trap}), this architecture prescribes qualitatively different behaviors depending on epistemic state. When the robot first detects anomalous drift, $\kappa$ spikes, the system cannot distinguish positional error from mechanical fault. The architecture responds by halting goal-directed navigation and initiating diagnostic maneuvers: deliberate weight shifts or diagnostic leg swings designed to isolate which wheel is loose. These actions sacrifice immediate progress to satisfy the tightened safety budget $\delta(\kappa )$. As the fault hypothesis sharpens and $\kappa$ drops, the agent transitions through a cautious navigation phase before eventually resuming efficient locomotion with full confidence in both its position and its dynamics model.
Table~\ref{tab:regimes} summarizes these behaviors. A key benefit is interpretability. By tying safety directly to the measurable $\kappa$ metric, we can verify that the robot acts conservatively exactly when it lacks the knowledge to operate safely.

Learning the optimal modulation function $\delta(\kappa)$ remains an open 
challenge. Hand-crafting it risks being either too conservative (sacrificing 
performance unnecessarily) or too permissive (allowing unsafe actions during 
high-uncertainty transients). Meta-learning across distributions of dynamics 
shifts offers a principled alternative, enabling the system 
to discover the optimal performance-safety trade-off and achieve ``anytime 
safety'' that holds throughout adaptation.

\section{Toward Tractable Implementation}

While this adaptive safety framework provides a principled conceptual foundation, 
practical deployment requires progress on several fronts: tractably approximating the 
joint belief $b(s,\theta)$, reducing the dimensionality of the parameter space through 
structural priors, leveraging foundation models for initialization and world modeling, 
and developing benchmarks that rigorously test compound uncertainty. We address each in turn.

\subsection{Approximating Joint Beliefs}\label{sec:approx_beliefs}

Implementing the Bayes-Adaptive framework ($P_{\theta}$) requires maintaining 
a joint belief $b_t(s, \theta)$. Exact inference is computationally intractable 
in continuous systems: state and dynamics uncertainties yield multi-modal, 
non-Gaussian posteriors that standard filters such as the Extended Kalman Filter 
cannot represent, as they collapse the distinct hypotheses illustrated in 
Figure~\ref{fig:epistemic_trap} into a single averaged estimate.

The tractable $\kappa$ proxy from Section~\ref{sec:approx_tractable} partially 
addresses this, but surfaces a key limitation for full deployment: the frozen 
ensemble cannot update from information-seeking probe responses, so the feedback 
loop cannot close at inference time. A practical resolution is a dual-ensemble 
design, one frozen ensemble anchoring the baseline prior for stable $\sigma_\theta$ 
measurement, and one online-updating ensemble trained from probe transitions to 
track shifted dynamics. This preserves measurement stability while restoring the 
information-seeking feedback loop.

A related limitation concerns temporal uncertainty. Our empirical results (Section~\ref{sec:emp_cost}) show that action delays roughly doubled the rate of meaningful super-additive failure, from roughly 9\% of configurations without delay to roughly 23\% of delay-related configurations, suggesting that when the causal link between action and observation is time lagged, standard Markovian beliefs collapse.

To address 
this, practical implementations should integrate Structured State-Space Models 
(SSMs) or Linear Recurrent Units (LRUs)~\cite{orvieto2023resurrecting}. Unlike 
standard RNNs which struggle with vanishing gradients over long horizons, these 
architectures maintain stable long-term memory traces, allowing the agent to 
correctly credit-assign errors to past actions even under variable delays.

\subsection{Structural Priors for Dimensionality Reduction}

Even with scalable approximations, the primary impediment to real-world deployment 
is the sheer dimensionality of the parameter space $\theta$. In high-stakes domains 
like digital medicine, this problem intensifies. An ICU sepsis management policy, 
for example, must operate in a high-dimensional observation space ($o$), monitoring 
over 100 continuously varying physiological signals such as heart rate and lactic 
acid. Crucially, we must distinguish these observations from the hidden state ($s$) 
(the patient's true pathological status, e.g., infection stage) which evolves 
according to hidden dynamics parameters ($\theta$), such as genetic drug resistance 
or metabolic rates. This yields a massive, partially observed joint space $(s, \theta)$. 
The challenge is that these unknown, patient-specific physiological ``rules'' ($\theta$) 
drive the evolution of the state, creating the Epistemic Trap, a decline in observed 
vitals ($o$) could stem from disease progression (state decay) or treatment inefficacy 
(dynamics mismatch). Without causal abstraction to isolate the parameters that 
specifically matter (e.g., renal function for a dosing decision), maintaining the 
full joint belief $b(s, \theta)$ becomes intractable.

Causal reinforcement learning~\cite{scholkopf2021toward} offers a necessary structural 
prior to overcome this. By representing the environment as a Structural Causal Model 
(SCM), we can factorize the belief state based on causal invariance. Causal abstraction 
identifies invariant mechanisms, whether the laws of kinematics in robotics or 
homeostatic baselines in physiology~\cite{scholkopf2022causality} and separates 
them from specific confounders (e.g., a metabolic disorder) that vary across 
deployments. This is critical for adaptive safety, the agent does not waste exploration 
budget probing invariant properties; instead, the belief need only track the subset of 
parameters causally relevant to the current prediction error, focusing DIS on targeted 
causal discovery rather than brute-force system identification. Actions become focused 
interventions designed to test specific conditional dependencies, effectively isolating 
the ``root cause'' of the epistemic coupling~\cite{pearl2009causality}. Approaches 
learning such invariant features have demonstrated robust generalization in adversarial 
settings~\cite{arjovsky2019invariant, krueger2021out}. By incorporating these priors, 
we reduce the effective dimensionality of the epistemic trap, making the calculation 
of $\kappa$ and the subsequent active resolution of uncertainty computationally feasible 
in real time.

\subsection{Foundation Models: Promising Initialization}

While causal priors reduce the search space, the agent still faces the \emph{cold 
start} problem: upon entering a new environment, the agent has almost no information about the current dynamics, leaving its joint belief spread thinly across a vast hypothesis space and making early decisions especially unsafe. Emerging foundation models (FMs) trained on diverse datasets of 
physical simulation and real-world data~\cite{zitkovich2023rt} represent a promising 
avenue to address this challenge. We identify three complementary roles for FM integration:

\textit{Initialization:}
FMs can dramatically reduce the initial entropy at deployment. Recent work provides 
concrete validation. Phys2Real~\cite{wang2025phys2real} demonstrates that vision 
language models can infer physical parameters (e.g., friction and mass) from visual 
observations alone, achieving high success on manipulation tasks where standard methods 
fail. To illustrate the breadth of this impact, consider precision agriculture: a 
standard irrigation controller must water a field for days to learn its specific soil 
drainage rates $\theta$, whereas a multimodal FM trained on global satellite data can 
infer these dynamics from visual cues such as soil color and vegetation texture, 
narrowing the hypothesis space and directly reducing initial $\kappa$ before a single 
drop of water is dispensed.

\textit{World Model Backbone:} Uncertainty-aware world models enable real-time 
computation of $\kappa$ by separating epistemic from aleatoric uncertainty. Recent advances in 
model-based RL demonstrate that explicitly decomposing these uncertainty types is essential to achieving strong performance in uncertain environments 
\cite{vlastelica2023mind, hansen2024tdmpc}.
Modern 
multimodal FMs could serve as the backbone of these world models, providing rich state 
representations that capture both semantic context (e.g., ``wilted crop'' or ``icy 
road'') and physical predictions, with safety penalties derived directly from the 
model's internal confidence.

\textit{Sample-Efficient Refinement:} Strong FM priors make DIS exponentially more 
efficient, shifting it from blind parameter-space exploration to targeted validation of 
FM predictions where uncertainty remains high. The convergence of scale, multimodality, 
and explicit uncertainty quantification makes this symbiosis of large-scale priors and 
active local refinement a particularly viable path to safe, general-purpose autonomy.

\subsection{Benchmarking Compound Uncertainty}

Rigorous validation requires benchmarking infrastructure that directly tests the 
\textit{interaction} of uncertainties, not merely their co-occurrence. Several 
benchmarks can technically combine partial observability with dynamics variation. 
The Real-World RL Suite~\cite{dulac2021challenges} supports simultaneous challenges
including system delays and sensor noise. Similarly, CARL~\cite{benjamins2022contextualize} enables hidden 
context settings where agents must infer changing physics parameters (e.g., gravity, 
friction) without direct observation. More recently, Gym4ReaL~\cite{salaorni2025gym4real} 
and POBAX~\cite{tao2025benchmarking} have expanded coverage of real-world challenges 
and partial observability forms, respectively.

However, these benchmarks treat perturbation sources as \textit{independent} rather 
than \textit{coupled}, providing no mechanism to measure interaction effects. 
POPGym~\cite{morad2023popgym} offers rigorous partial observability evaluation but 
mandates stationary dynamics within episodes. Robust-Gymnasium~\cite{gu2025robust} 
provides modular disruptions across observations, actions, and dynamics, yet evaluates 
robustness to each perturbation type separately. None provide explicit metrics for 
epistemic coupling ($\kappa$), tasks engineered to maximize state dynamics ambiguity, 
or evaluation protocols that distinguish super-additive failures from additive ones.

We propose the development of \textit{POMDP-Robust-Gym}, a benchmark explicitly 
designed for compound uncertainty evaluation. To move beyond ``passive robustness,'' 
this suite requires three critical features:

\begin{enumerate}
    \item \textbf{Combined Failure Modes with Engineered Ambiguity:} Scenarios must 
    pair specific sensor occlusions with dynamics shifts (e.g., hidden joint sensors + 
    delayed actuation) such that the resulting observations are consistent with multiple 
    incompatible hypotheses about the world state. The design goal is to maximize 
    epistemic coupling ($\kappa$), creating conditions where standard methods cannot 
    isolate the source of prediction error.

    \item \textbf{Epistemic Observability for Evaluators:} The infrastructure must 
    expose the ground-truth dynamics parameters to the evaluator (but not the agent), 
    enabling real-time calculation of an \textit{Epistemic Gap}, measuring the 
    divergence between the agent's belief $b(s,\theta)$ and the true state dynamics 
    pair, thereby allowing us to quantify \textit{when} an agent is confused, rather than 
    merely observing \textit{that} it failed. This mirrors CARL's ~\cite{benjamins2022contextualize}  provision of ground-truth context but extends it to joint state-dynamics tracking.

    \item \textbf{Information-Aware Metrics:} Standard cumulative reward is a 
    ``greedy'' metric that penalizes the cautious exploration required for safety. 
    \textit{Regret-based metrics} that reward timely uncertainty reduction are 
    needed: an agent that sacrifices 5\% of its speed to verify a safety constraint 
    should score \textit{higher} than a reckless agent that succeeded by chance. 
    Such metrics would complement existing robustness evaluations~\cite{dulac2021challenges} 
    with explicit credit for epistemic prudence.
\end{enumerate}

Our Walker2D study serves as a template for this methodology: the systematic variation across 120 configurations demonstrated not only that super-additive failure is detectable, but that elevated $\kappa$ tracks the regimes in which such failure occurs, with $\kappa$ spikes preceding the observed velocity collapses in our representative run. This positions $\kappa$ as a candidate early warning signal, to be validated at scale in future benchmarks rather than treated as a post-hoc diagnostic.
Future benchmarks must extend this methodology by instrumenting environments to detect 
DIS behavior automatically. By tracking whether an agent executes diagnostic maneuvers 
when uncertainty rises, we can rigorously verify the core thesis of this perspective: 
that true safety in open worlds requires active information management, not just 
worst case bounding.

\section{Conclusion and Broader Implications}

The fundamental barrier to deployable, safety-critical reinforcement learning is not 
dynamics shift itself, nor partial observability in isolation, but the \textit{Epistemic 
Trap} created by their compounding interaction. By reframing safety from a static 
constraint to a dynamic variable managed through information quality, we can break 
this cycle. Together, $\kappa$ based uncertainty quantification and active perception via dynamics information seeking form the core of this architecture, with causal structural priors and foundation model integration offering a tractable path toward deployment at scale.

Looking ahead, this information driven approach promises broader impact beyond 
robotics. Across domains, from diagnostic AI that actively requests targeted tests 
when patient observations are insufficient, to autonomous inspection systems that seek 
better vantage points under degraded sensing, the common thread is treating knowledge 
gaps not as unfortunate realities to be bounded, but as manageable quantities to be 
actively resolved.

The goal is AI systems that are not merely conservatively robust, but intelligently adaptive: recognizing their own ignorance, seeking information strategically, and modulating behavior based on what they know and how confidently they know it. Achieving this vision requires community investment in the 
mathematical foundations (information-theoretic frameworks,  and objectives), 
architectural innovations (uncertainty-aware world models, causal abstraction), and 
empirical infrastructure (POMDP-Robust-Gym) outlined here. This vision becomes 
tractable through foundation models: web-scale pretraining can provide the robust 
dynamics priors our framework requires, and the next generation of deployable safe AI 
will emerge from the combination of foundation model priors and active perception 
refinement, a path equally essential for clinical decision support, autonomous 
vehicles, and other safety-critical domains.

\bibliography{reference}

\begin{appendices}
 
\section{Information-Theoretic Grounding for the Tractable Approximation}
\label{app:kappa_derivation}
 
We show that $\sigma_\theta + \sigma_s$ constitutes a provable, monotone upper bound on the conditional mutual information $I(s;\theta \mid b_t)$. Under the given assumptions, this provides a theoretically grounded and tractable proxy for online computation.
 
\paragraph{Upper bound.}
Applying the chain rule of entropy to the joint belief $b_t(s,\theta)$:
\begin{equation}
    I(s;\,\theta \mid b_t)
    \;=\; H(s \mid b_t) \;+\; H(\theta \mid b_t) \;-\; H(s,\theta \mid b_t).
    \label{eq:app_chain}
\end{equation}
Since joint entropy is non-negative, $H(s,\theta \mid b_t) \geq 0$, it follows
immediately that:
\begin{equation}
    \boxed{I(s;\,\theta \mid b_t) \;\leq\; H(s \mid b_t) \;+\; H(\theta \mid b_t).}
    \label{eq:app_ub}
\end{equation}
 
\paragraph{Proxy identification.}
The normalised ensemble MSE z-score ($\sigma_{\theta}$) is a proxy for $H(\theta \mid b_t)$. Under the bootstrapped ensemble design, the ensemble prediction variance is a consistent estimator of epistemic uncertainty over $\theta$. Assuming isotropic Gaussian posteriors, under which $\text{tr}(\hat{\Sigma}_{\theta}) = M \cdot \text{MSE}_t$, the Gaussian differential entropy $H(\theta \mid b_t) = \frac{1}{2} \ln \det(2\pi e \hat{\Sigma}_{\theta})$ is monotonically increasing in $\text{tr}(\hat{\Sigma}_{\theta})$, and hence in $\text{MSE}_t$. The normalized z-score $\sigma_{\theta}$ therefore preserves this monotone ordering, serving as a tractable proxy for $H(\theta \mid b_t)$ without requiring explicit posterior inference.
Similarly, the state observability deficit ($\sigma_s$) serves as a proxy for
$H(s \mid b_t)$: masked observation dimensions directly inflate state entropy
by removing information from the posterior, and action delays sever the
feedback channel that would otherwise reduce it.
Together:
\begin{equation}
I(s;\,\theta \mid b_t)
    \;\leq\; H(s \mid b_t) \;+\;
H(\theta \mid b_t)
    \;\propto\; \sigma_\theta \;+\; \sigma_s \;=\; \kappa.
    \label{eq:app_chain_full}
\end{equation}
 
\paragraph{Why this suffices.}
The approximation $\kappa  \approx \sigma_\theta + \sigma_s$ uses
this upper bound as a tractable surrogate.
While the bound is not tight in general, the slack equals $H(s \mid \theta, b_t)$ the residual state uncertainty that remains even after dynamics are known. This slack does not vanish in the high-coupling regime; the proxy $\sigma_\theta + \sigma_s$ therefore overestimates $I(s; \theta \mid b_t)$ throughout. However, the property required for reliable regime detection is not tightness but monotone co-variation: as epistemic coupling rises, both $H(s \mid b_t)$ and $H(\theta \mid b_t)$ increase, and so does their sum. 

Conversely, when uncertainties decouple ($\kappa  \rightarrow 0$), both marginal entropies shrink, pulling the proxy toward zero. The proxy therefore correctly orders regimes, small in the decoupled case, large in the Epistemic Trap,without requiring the bound to be tight. This is analogous to the standard Bayesian variance decomposition, where total predictive variance is decomposed into aleatoric and epistemic terms via the law of total variance: exact only under Gaussian assumptions, but monotone and empirically reliable for downstream use ~\cite{lakshminarayanan2017simple}
The proxy therefore captures the proportional scaling of epistemic coupling
without requiring full joint-belief inference.

\section*{Appendix B\quad Calibration of Regime Thresholds}
\label{app:threshold_calibration}

The regime boundaries $\tau_{\text{low}}$ and $\tau_{\text{high}}$ are
environment-specific hyperparameters that partition the proxy
$\kappa \approx \sigma_\theta + \sigma_s$ into three operationally
distinct zones.
For the Walker2D proof-of-concept, they are derived in two steps.

\paragraph{Step 1: Anchor the baseline noise floor.}
The bootstrapped ensemble is pre-trained on $T_{\text{pre}} = 300$
unperturbed transitions from the C1 (standard) environment, recording
the mean $\mu_0$ and standard deviation $\sigma_0$ of the ensemble MSE.
This establishes a stable prior for normal locomotion kinematics and
fixes the denominator of the $\sigma_\theta$ z-score
(Eq.~\ref{eq:MSE_ensemble} of the main text).
Under unperturbed conditions the proxy evaluates to
$\kappa(C_1) = 0.036$ (post-onset mean, seed~1024), reflecting
residual aleatoric noise only.
The 95th percentile of the C1 $\kappa$ distribution is approximately
$0.15$; $\tau_{\text{low}} = 0.2$ is set just above this value,
providing a conservative margin that ensures normal operation is
classified as Low Deficit even under stochastic fluctuations.

\paragraph{Step 2: Bracket the individual-stressor regimes.}
With $\tau_{\text{low}}$ fixed, $\tau_{\text{high}}$ is set by
inspecting the proxy values under the two isolated stressors:

\begin{center}
\begin{tabular}{lcc}
\hline
Scenario & Post-onset $\kappa$ mean & Assigned regime \\
\hline
C1 (standard)       & 0.036 & Low Deficit ($< \tau_{\text{low}}$) \\
C2 (masked sensors) & 0.089 & Low Deficit ($< \tau_{\text{low}}$) \\
C3 (action delay)   & 0.332 & Transition \\
C4 (combined)       & 0.496 & Transition / High Deficit boundary \\
\hline
\end{tabular}
\end{center}

The threshold $\tau_{\text{high}} = 0.5$ is set at the natural gap between the single-stressor regime (C3, mean $0.332$) and the compound regime (C4, mean $0.496$). Validation is twofold: $\kappa$ spikes in C4 peak at $1.376$ ($t \approx 350,704$), correctly triggering High Deficit classification immediately before the forward-velocity collapses in Figure~\ref{fig:walker_results}; and the C4 interaction effect is statistically significant ($p < 10^{-4}$, $413.19$ additional loss units beyond the additive prediction), confirming the threshold demarcates the regime where passive observation fails.

\paragraph{Generalization to other domains.}
The two-step procedure is domain-agnostic.
$\tau_{\text{low}}$ should be set above the 95th percentile of
$\kappa$ recorded during an unperturbed baseline rollout;
$\tau_{\text{high}}$ should be set at the upper tail of $\kappa$
observed under the most demanding \emph{individual} stressor
available in the target domain.
Both thresholds reflect the operational risk tolerance of the
deployment environment and should be validated against known-safe
and known-unsafe conditions before deployment.
\end{appendices}

\end{document}